\documentstyle[12pt]{article}
      \newcommand{\beq}{\begin{equation}}
      \newcommand{\eeq}{\end{equation}}
      \newcommand{\beqa}{\begin{eqnarray}}
      \newcommand{\eeqa}{\end{eqnarray}}
      
      \newcommand{\nn}{\nonumber}
      \newcommand{\Tr}{{\rm Tr}}
      
      \newcommand{\bra}{\left\langle}
      \newcommand{\ket}{\right\rangle}
      
      \newcommand{\del}{\partial}

      \newcommand{\al}{\alpha}
      \newcommand{\be}{\beta}
      
      \newcommand{\de}{\delta}
      \newcommand{\ep}{\epsilon}
      
      \newcommand{\ka}{\kappa}
      \newcommand{\la}{\lambda}

      \newcommand{\bz}{\mbox{\boldmath $z$}}
      \newcommand{\const}{{\rm const. \, }}
      \newcommand{\trans}{\, {}^t \!}
\begin{document}
\begin{titlepage}
%
\begin{center}
{\large{\bf REPLICA METHOD FOR WIDE CORRELATORS
IN GAUSSIAN ORTHOGONAL, UNITARY  AND SYMPLECTIC RANDOM MATRIX
ENSEMBLES}} \\[12mm]
%
%
 $\mbox{{\sc Chigak Itoi}}^{\, a, }$
\footnote{\tt e-mail: itoi@phys.cst.nihon-u.ac.jp} \\[3mm]
 $\mbox{{\sc Hisamitsu Mukaida}}^{\, b, }$
\footnote{\tt e-mail: mukaida@thaxp1.ins.u-tokyo.ac.jp} \\[2mm]
{\center and}\\[2mm]
 $\mbox{{\sc Yoshinori Sakamoto}}^{\, a, }$
\footnote{\tt e-mail: yossi@phys.cst.nihon-u.ac.jp} \\[5mm]
${}^a$ {\it Department of Physics, 
College of Science and Technology,  \\
Nihon University, Kanda Surugadai, Chiyoda-ku, 
Tokyo 101, Japan} \\[3mm]
${}^b$ {\it Department of Physics, Saitama Medical College, }\\
{\it Kawakado, Moroyama, Saitama, 350-91, Japan} \\[3mm]
{\bf Abstract}\\[5mm]
{\parbox{13.5cm}{\hspace{5mm}
We calculate connected correlators in Gaussian orthogonal,
unitary and symplectic random matrix ensembles by the replica method 
in the $1/N$-expansion.  
We obtain averaged one-point Green's functions up to the next-to-leading 
order $O(1/N)$, wide two-level correlators up to the first nontrivial order
 $O(1/N^2)$ and wide three-level correlators up to the first 
nontrivial order $O(1/N^4)$ by carefully treating fluctuations in saddle-point 
evaluation.\\[3mm]
}}
\end{center}
\vspace*{5ex}
\end{titlepage}

\section{Introduction}
Recently, the universality of the   
wide distance two level connected correlator in random matrix theories
is shown by Ambj{\o}rn, Jurkiewicz and Makeenko \cite{AJM}.
Br{\'e}zin and Zee regarded this universality as importance 
from the viewpoint of physics in disordered systems \cite{BZ}. 
Since the correlator 
does not depend on the probability distribution of a matrix ensemble,
we may use the correlator calculated in the 
corresponding simple Gaussian ensemble with the same symmetry
for the non-trivial disordered system in concern. 
Compared to the short distance correlator which is already
well-known as a universal quantity \cite{mehta},   
the wide correlators are able to be calculated explicitly 
in extensive types of ensembles
in various ways \cite{B,BHZ}.
Actually, one observes the strongly universal properties  
which random matrix theories have themselves in the wide connected correlators.
Their universality classification is also done in \cite{AA}. 
These mathematical studies of random matrix 
theories themselves enable us to recognize 
the real universal nature of level statistics.

In this paper, we    
examine the replica method to calculate 
the wide connected two level correlator 
in gaussian orthogonal (GOE), unitary (GUE) and symplectic (GSE) ensembles.
These simple ensembles GOE, GUE and GSE can 
describe a time reversal system without spin, 
a general system and a time reversal system with spin, respectively 
\cite{mehta}. 
The replica method is well-known as a convenient scheme to
calculate two body Green's function in some models of 
the Anderson localization \cite{MS}. 
Even though this method cannot work well
to calculate the short distance correlator \cite{VZ},
here we show this method is useful to calculate  the
wide two level correlator in the $1/N$ expansion.
This method is much simpler 
than the supersymmetry method which 
enables us to calculate both the wide and short distance correlators 
\cite{VZ,Guhr}.
Here, we remark the reason why the replica method is not 
good for the short distance correlator. 
We are going to calculate the explicit form of the two-level and three-level
correlators in GOE, GUE and GSE, which are identical to those 
calculated by solving functional equations \cite{B,I} or by diagrammatic
method \cite{IS}. 
Here we study a matrix theory defined 
by the following probability distribution 
\beq
        P(H) = \frac{1}{Z_H}  \exp \left( -\frac{N}{2} \Tr H^2 \right), \ \ 
        {Z_H} \equiv \int D H \, \exp\left( -\frac{N}{2} \Tr H^2 \right),
\label{prob} 
\eeq
where $H$ is an $N \times N$ matrix. 
The explicit form of the measure $DH$ depends on 
a type of an ensemble of $H$.  
Our interest is focused on computing averaged Green's functions 
by the replica method. We show their calculation method explicitly 
for GOE in section 2, for GUE in section 3 and GSE in
section 4.
 
\section{Gaussian orthogonal ensemble}
To begin with, we explore the case of the GOE which is defined by 
the ensemble of real symmetric matrices  obeying the  probability distribution (\ref{prob}).  
The measure $D H$ is explicitly written as 
\beq
        D H = \prod_{k = 1}^N dH_{k\, k} \prod_{i<j} dH_{i\,j}.
\label{omeasure}  
\eeq
\subsection{One-point function}
We calculate the averaged one-point function 
\beq
        G(z) \equiv \langle \frac{1}{N} \Tr \frac{1}{z - H - i \ep} \rangle
         \equiv \int DH \,P(H) 
        \frac{1}{N}\Tr \frac{1}{z - H - i \ep}. 
\label{G(z)}  
\eeq
Note that $ -i \, \ep$ $(\ep > 0)$ means that $G(z)$ considered here
 is the advanced Green's  function.  We assume 
\beq
        |z| >  \sqrt{2}
\label{assumption}
\eeq
in the case of the GOE. The meaning of this assumption will be 
clarified later.  
Results for $|z| < \sqrt{2}$ 
will be extracted employing the analytic continuation that is 
uniquely determined by $ - i \ep$.  

In order to apply the replica method, we first introduce 
the $n$-flavor real vectors  having $N$ component 
$(\phi^{a}_{i})_{1 \leq i \leq N}$, where the superscript $a$, 
which runs from $1$ to $n$,  specifies the flavor.   
Let us consider the following integration over $\phi$ for 
constructing a generating function of $G(z)$:   
\beq
 \Gamma [z, H] \equiv \frac{1}{Z_\phi} \, 
 \int D  \phi \, \exp \{ -\frac{i}{2}  \trans \phi^a \, 
 (z - H - i \ep) \, \phi^a  \}, \ \
 Z_\phi = \int D  \phi \, \exp \left( - \frac{1}{2} 
 \trans \phi^a \phi^a \right),
\eeq
where $\trans \, $  means transposition and 
the summation over the repeated indices are implicitly taken. 
The measure is defined by 
\beq
D \phi \equiv \prod_{a=1}^n\prod_{i=1}^N d \phi^{a}_i.  
\eeq
Note that the integration over 
$\phi$ for $\Gamma$ converges owing to $-i\ep$.  
Hereafter we do not explicitly write $- i \ep$ for brevity. 

Taking the derivative with respect to $z$, we have 
\beqa
        \frac{\del \Gamma [z, H]}{\del z} &=& \frac{\del}{\del z} 
        {\rm det}^{-n/2}( i (z - H))  \nn\\
        &=& - \frac{n}{2} \Tr \frac{1}{z - H}  \, 
        {\rm det}^{-n/2}( i (z - H)). 
\label{G(z,H)}
\eeqa
We define  $W(z)$ as the average of $\Gamma$ over  $H$: 
\beq
        W(z) \equiv  \langle  \Gamma [z,H] \rangle. 
\label{W}
\eeq
It follows from eq.~(\ref{G(z,H)}) that 
\beq
        G(z) = \lim_{n \rightarrow 0} \left(-\frac{2}{n N} \right) 
        \frac{d}{d z} W(z),  
\label{WtoG} 
\eeq
which indicates that $W(z)$ plays a role of a generating function of 
$G(z)$. 
Note that we need to take the limit $n \rightarrow 0$ in order to eliminate 
the determinant factor in eq.~(\ref{G(z,H)}). 
 
Let us compute  $W(z)$. The average over $H$ in eq.~(\ref{W}) is 
explicitly carried out and we have 
\beq
        W(z) = \frac{1}{Z_\phi} \, \int D \phi \,  \exp 
        \left( -\frac{i}{2}  \trans \phi^a \, 
         z  \, \phi^a -\frac{1}{8 N} ( \trans \phi^{a} \phi^{b})^2  \right).  
\label{W(z)}
\eeq
The quartic interaction can be represented as 
\beq
        \exp \left( -\frac{1}{8 N} ( \trans \phi^{a} \phi^{b})^2 \right ) = 
        \frac{1}{Z_Q} \, \int D Q \exp \left( - \frac{N}{2} Q_{a \, b} 
        Q_{b \, a} + 
        \frac{i}{2} \trans \phi^a  \phi^b Q_{a\, b}\right),   
\label{aux}
\eeq
where $(Q)_{a b}$ is  the real symmetric $n \times n$ matrix and the 
normalization factor is defined by 
\beq
        Z_{Q} \equiv \int D Q \exp \left( - \frac{N}{2} Q_{a \, b} 
        Q_{b \, a} \right). 
\eeq
Using the representation (\ref{aux}) we get 
\beq
        W(z) =  \frac{1}{Z_\phi Z_Q}\, \int D Q D \phi \,  \exp \left( - 
        \frac{N}{2} \Tr \, Q^2 - 
        \frac{i}{2} \trans \phi_i (z - Q) \phi_i \right ). 
\label{W(z)byQ} 
\eeq
Note that indices $a$ and $b$ are implicitly summed in the above notation: 
we regard $\phi_i$ as an $n$-component vector.  

The $\phi$-integral becomes Gaussian and we find 
\beq
        W(z) = \frac{1}{Z_Q} \int D Q \, \exp \{ -\frac{N}{2} 
        \Tr ( \log i(z - Q) + Q^2 ) \}. 
\eeq
It can be written as integration over eigenvalues \cite{mehta}
\beqa
        W(z) =  \frac{C_n}{Z_Q} \int  \prod^{n}_{a=1} du_a \ 
        \left|  \Delta(u) \right| \  \exp \{ -N \sum^{n}_{a = 1} g(u_a) \},
\label{W(z)inu}
\eeqa
where $C_n$ is a constant that depends on $n$ and 
$\Delta(u)$ is the Van der Monde determinant 
\beq
        \Delta(u) \equiv \prod_{a < b} (u_a - u_b).   
\eeq
The function $g$ appeared in the exponent is defined by 
\beq
        g(x) \equiv \frac{1}{2} ( \log i (z - x) + x^2).
\eeq

Now we compute $W(z)$ by the $1/N$-expansion.  Let us first 
compute it in the leading order. 
Since the contribution from the Van der Monde determinant 
in eq.~(\ref{W(z)inu}) is neglected in the leading order, 
$u_a$'s are completely decoupled each other and 
$W(z)$ is evaluated by a single-variable integration.  i.e., 
\beq
        W(z) \sim  \const \left( \int du  \, \exp \left\{ -{N}  g(u) 
        \right\} \right)^n   
\label{leading}
\eeq 
in the leading order. 
We can evaluate  the right-hand side of (\ref{leading}) by the 
saddle-point method.  
The saddle-point equation is 
\beq
        g'(u) = - \frac{1}{2(z - u)} +  u = 0.
\label{sp}
\eeq
We have the two solutions $u_{\pm}$, where 
\beq
        u_{\pm} = \frac{1}{2}(z \pm \sqrt{z^2 - 2}). 
\label{upm} 
\eeq
We find that contribution to the integral (\ref{leading}) from 
$u \sim u_-$ dominates over that from $u \sim u_+$.  In fact, 
a straightforward calculation gives 
\beq
{\rm Re} \, g(u_-) <  {\rm Re} \, g(u_+),  \ \ {\rm if}  \ z > \sqrt{2}. 
\eeq 
Therefore we get 
\beq
        W(z) \sim \const e^{ - N n g(u_-) }
\label{leadingW}  
\eeq
in the leading order. 
Inserting the above result into eq.~(\ref{WtoG}), we obtain the 
well-known result 
\beq
        G(z) = z - \sqrt{z^2 - 2}, 
\eeq
where the overall constant in eq.~(\ref{leadingW}) is chosen such 
that $G(z)$ should satisfy 
 the boundary condition $G(z) \rightarrow 1/z$ as 
$z \rightarrow \infty$.

We can compute higher-order corrections by expanding $g(u_a)$ in 
eq.~(\ref{W(z)inu}) around $u = u_-$.   
\beq
        g(u_a) = g(u_-) + \frac{1}{2N} g''(u_-) \, y_{a}^2 + 
                        \sum_{n = 3}^{\infty} \frac{N^{-n/2} }{n!} g^{(n)}(u_-) 
\, y_{a}^n, 
\label{expandedg}  
\eeq
where we denote 
\beq
        u_a - u_- = N^{-1/2} y_a.  
\eeq

Here we shall calculate up to the first-order correction,  which is sufficient 
for determining the connected two-point function,  as we will see 
below. To this end, 
we can neglect the cubic and the higher terms in eq.~(\ref{expandedg}). 
Then we find  
\beq
        W(z) \sim C_n e^{ -N n g(u_-) } \, \int \prod_a d y_a \, 
        \left| \Delta (y) \right| 
        \exp \{-\frac{1}{2} \sum_{a = 1}^n g''(u_-) y_a^2\}   
\eeq
up to the next-to-leading order. 
Note that 
\beqa
        g''(u_-) &=& 1 - 2 u_-^2 \nn\\
                     &=& \sqrt{z^2 -2} \left(z - \sqrt{z^2 - 2} \right) , 
\label{g''}
\eeqa
which is real and positive under the assumption (\ref{assumption}).  
Therefore the integration over $y_a$ is easily performed going back to 
representation of an integration over real symmetric matrices.  
Since there are $n(n+1)/2$ 
independent variables in a real symmetric matrix, we get
\beqa
        \frac{C_n}{Z_Q} \, \int \prod_a d y_a \, \left| \Delta (y) \right| 
        \exp \{-\frac{1}{2} \sum_{a = 1}^n g''(u_-) y_a^2\} &=& 
        \frac{1}{Z_Q}\, \int Dq \exp \{-\frac{1}{2}  g''(u_-) \sum_{a,b} 
        q_{a\, b}^2\} \nn\\
        &=& \left( \frac{1}{g''(u_-)} \right)^{n(n+1)/4}. 
\label{saddle} 
\eeqa
Thus we conclude 
\beq
      W(z) \sim  e^{-N n g(u_-) } \left( \frac{1}{g''(u_-)} \right)^{n(n+1)/4}
\eeq
up to the next-to-leading order. 
Using eqs.~(\ref{WtoG}) and (\ref{g''}), 
we obtain $G(z)$ including the first-order correction: 
\beq
        G(z) = \left( z - \sqrt{z^2 - 2} \right) \left(1+\frac{1}{2 N} 
        \frac{1}{z^2 - 2} \right)
        + O(1/N^2).  
\eeq
This result is identical to that obtained in \cite{I}.

Here we comment on the validity of the saddle point method in 
eq.~(\ref{saddle}). The saddle point evaluation in the $1/N$ expansion 
holds only in the region where $g''(u_-)$ is $O(N^0)$. This approximation 
becomes incorrect for $z=\pm \sqrt{2}+ O(1/N)$ 
which yields $g''(u_-) = O(1/N)$.

\subsection{Two-point function}
Now we turn to the two-point function 
\beq
        G(z_1, \,  z_2) \equiv \bra \frac{1}{N} \Tr \frac{1}{z_1 - H} \, 
        \frac{1}{N} \Tr \frac{1}{z_2 - H} \ket. 
\label{G2} 
\eeq
We shall show how to compute {\it wide} correlation by the 
replica method.  i.e.
, we 
assume $z_1 - z_2 \sim O(N^0)$ as well as the condition 
(\ref{assumption}). 

As is the case of the one-point function, we start with the following integral
\beq
        \Gamma [z_1, z_2, H] \equiv \frac{1}{Z_{\phi_1}Z_{\phi_2}} \int D\phi_1 
D\phi_2 \, 
         \exp \{ -\frac{i}{2} \,   \trans \phi^a_{1} \, (z_1 \,  - H) \, \phi^a_
{1} 
                        -\frac{i}{2} \,  \trans \phi^p_{2} \, (z_2 \,  - H ) \, 
\phi^p_{2}  \}.
\eeq
Here we have introduced the two species of vectors labeled by 1 and 2.  
The two species respectively have $n$- and $m$- flavor, which means that  
$ 1 \leq a \leq n $ and that  $1 \leq p \leq m$.
We average $Z[z_1, z_2, H]$ over $H$ and define $W(z_1, z_2)$,  
a generating function of $G(z_1, \,  z_2)$. 
\beq
        W(z_1, \, z_2) \equiv \bra \Gamma [z_1, \, z_2, H] \ket.  
\eeq

The two-point function is derived from $W(z_1, z_2)$ in the 
following formula: 
\beq
        G(z_1, \, z_2) = \lim_{n, m \rightarrow 0} 
        \left( \frac{-2}{Nm}\right) \left(\frac{-2}{Nn}\right) 
\frac{\partial ^2}{\partial z_1 \partial z_2} W(z_1, \, z_2).
\label{W2toG2}
\eeq

Our strategy for computing $W(z_1, \, z_2)$ is the same as in the case
 of the one-point function.  Namely, we first perform 
the integration over $H$ of $W(z_1, z_2)$.  Second, we introduce the 
auxiliary matrix $Q$ in order to carry out the integration over 
$\phi_1$ and $\phi_2$.  The integration over the vector variables 
bring us the theory described by $Q$, which can be 
 investigated by the saddle-point method for large $N$. 

The first step is easily performed.  We have 
\beq
        W(z_1, \, z_2) = \frac{1}{Z_{\phi_1}Z_{\phi_2}} 
        \int D\phi_1 D\phi_2 \, 
         \exp \{ -\frac{i}{2} \,   \left( \trans \phi^a_{1} \, z_1 \, 
         \phi^a_{1} + 
         \trans \phi^p_{2} \, z_2 \, \phi^p_{2}  \right)   -  
         \frac{1}{8 N} \left( \trans \phi_{1}^a \, \phi_{1}^b + 
        \trans \phi_{2}^p \, \phi_{2 }^q \right)^2 \}. 
\label{W2}
\eeq
Next, let us introduce the $(n + m) \times (n + m) $ real 
symmetric matrix $Q$ using the following expression:
\beq
        Q = 
        \left(
        \begin{array}{cc}
                Q^{1  1} & Q^{1 2} \\
                Q^{2 1} & Q^{2 2} \\
        \end{array}
        \right), 
\eeq 
where $ Q^{1  1} $ and $ Q^{2  2} $ are $n \times n$ and $m \times m$ 
real symmetric matrix respectively. 
Since $Q$ is real-symmetric,  $Q^{1 2}$,  which is an $n \times m$
 matrix,  satisfies 
$Q^{1 2} = {}^t Q^{2 1}$.  In order to  write down a formula
 corresponding to (\ref{W(z)byQ}), we employ the following notations: 
\beq
        \Phi_i \equiv \left(
       \begin{array}{c}
	   \phi_{1 i} \\
       \phi_{2 i}
       \end{array}
                             \right),  
\eeq 
which is $(n + m)$-component vectors, and 
\beq
        \bz  = 
        \left(
        \begin{array}{cc}
                z_1 I_n & 0 \\
                0 & z_2 I_m \\
        \end{array}
        \right),  
\eeq
where $I_n$ is the unit $n \times n$ matrix.  We can check that the
 right-hand side of eq.~(\ref{W2}) is identical with 
\beq
         \frac{1}{Z_Q Z_{\phi_1}Z_{\phi_2}}\int D Q D\phi_1 D\phi_2 \, 
         \exp \{ - \frac{N}{2} \Tr \, Q^2 - \frac{i}{2} \, {}^t \Phi_i \, 
        (\bz - Q) \, \Phi_i \}. 
\eeq
After integrating the vector variables, we acquire
\beq
        W(z_1, \, z_2) = \frac{1}{Z_Q} \int DQ \, 
        \exp \left \{-\frac{N}{2} \Tr  \left(\log i(\bz - Q) +  Q^2 \right) 
        \right \}. 
\label{W2byQ} 
\eeq

We analyze the above integral formula by the saddle-point method. 
The saddle-point equation is 
\beq
        (\bz - Q)^{-1} = 2 Q.
\label{sp2} 
\eeq
Or equivalently, 
\beqa
        2 Q^{1 1}(z_1 - Q^{1 1}) - 2 Q^{1 2} Q^{2  1} &=& 1 \nn\\
        - 2 Q^{1 1} Q^{1  2} + 2 Q^{1 2}(z_2 - Q ^{2  2}) &=& 0 \nn\\
        2 Q^{2 1}(z_1 - Q^{1  1} ) - 2 Q^{2  2} Q^{2 1} &=& 0 \nn\\
        - 2 Q^{2 1} Q^{1  2} + 2 Q^{2 2} (z_2 - Q^{2  2}) &=& 1.  
\eeqa
After transposing the third equation and subtracting it from the second
 equation, we find 
\beq
        Q^{1  2} = Q^{2 1} = 0. 
\label{nondiagonal} 
\eeq
The remaining equations containing $Q^{1 1}$ and $Q^{2 2}$ are solved by
 diagonalization. 
Suppose that $Q^{1 1}$ and $Q^{2 2}$ are respectively diagonalized by
 $O_1$ and $O_2$.  Explicitly,  
\beq
        {}^t O_1 \, Q^{1 1} \, O_1  = {\rm diag \, } (u_1, \cdots, u_n),  \qquad
        {}^t O_2 \, Q^{2 2} \, O_2 =  {\rm diag \, } (v_1, \cdots, v_m). 
\label{transf} 
\eeq
Then the equations to be solved are 
\beqa
                 u_a  &=& \frac{1}{2(z_1 - u_a )},  \nn\\
                 v_p  &=& \frac{1}{2(z_2 - v_p )},
\label{remaining}
\eeqa
which are the same form as the saddle-point equation appeared in 
the 
previous subsection.  Following the argument on the most dominant 
saddle-point that is carried out in the one-point function, 
we should choose the saddle point 
\beqa
        u_a &=& u_-,  \qquad {\rm for \ all} \ a, \nn\\
        v_p &=& v_-,  \qquad {\rm for \ all} \ p.  
\label{diagonal}  
\eeqa
 Performing the inverse transformation of 
eq.~(\ref{transf}), we obtain the most dominant saddle point  
$\bar Q$ in the original basis:
\beq
        \bar Q = 
        \left(
        \begin{array}{cc}
                u_- I_n & 0 \\
                0 & v_- I_m \\
        \end{array}
        \right).   
\label{barQ}
\eeq

Next we consider fluctuations around the saddle point $\bar Q$. 
Using the identity 
\beq
        \Tr \log(A + \de A) = \Tr \log A + \Tr A^{-1} \de A - \frac{1}{2} 
        \Tr A^{-1} \de A \, A^{-1} \de A 
        + \cdots,
\label{expandlog} 
\eeq
and the saddle point equation (\ref{sp2}), we have
\beq
        \Tr \left( \log i(\bz - \bar Q-\frac{\de Q}{\sqrt{N}})  + \, 
        (\bar Q+ \frac{\de Q}{\sqrt{N}})^2  \right) =  
        \Tr \left( \log i(\bz - \bar Q) + \bar Q^2  \right)  + 
        \frac{1}{N} \Tr \, 
        ( \de Q^2 - 2 \de Q \, \bar Q \, \de Q \, \bar Q )  + \cdots. 
\eeq
Inserting the explicit form (\ref{barQ}) into the above, we see that the 
generating function becomes 
\beqa
        W(z_1, z_2) &\sim& \frac{1}{Z_Q} e^{-N n g(u_-)} \int D(\de Q^{11}) \, 
        \exp\{-\frac{1}{2} (1 - 2 {u_-}^2) \Tr \left( \de Q^{11} \right)^2 \} 
        \times \nn\\
        &{  }& e^{-N m g(v_-)} \int D(\de Q^{22}) \, 
        \exp\{-\frac{1}{2} (1-2{v_-}^2) \Tr \left( \de Q^{22} \right)^2 \}
        \times \nn\\
        &{  }& \int D(\de Q^{12}) \, \exp\{- (1 - 2 u_- v_-)
        \Tr \left( \de Q^{12} \right)^2 \}
\eeqa
up to the next-to-leading order. 
Remembering that $g''(u_-) = 1 - 2 u_-^2$,  we get 
\beqa
        W(z_1, z_2) &=& e^{ -N n g(u_-)} 
        \left( \frac{1}{g''(u_-)} \right)^{n(n+1)/4} 
        e^{ -N m g(v_-)} \left( \frac{1}{g''(v_-)} \right)^{m(m+1)/4} 
        \left( \frac{1}{1 - 2 u_- v_-} \right)^{nm/2} \nn\\
        &=& W(z_1) W(z_2) \left( \frac{1}{1 - 2 u_- v_-} \right)^{nm/2}.
\label{W2-2}
\eeqa
Inserting the above formula into  eq.~(\ref{W2toG2}),  we can derive the
 two-point function. 
 Differentiating $W(z_1)$ and $W(z_2)$ brings about the disconnected
 part, $G(z_1) G(z_2)$, 
 while the last factor contributes to the connected part. 
Thus we obtain 
\beq
        G(z_1, z_2) = G(z_1) G(z_2) - \frac{2}{N^2} 
        \frac{\del^2}{\del z_1 \del z_2} \log \left( 1 - \frac{1}{2} G(z_1) 
        G(z_2) \right),  
\label{GOE2}
\eeq
where we have used eq.~(\ref{WtoG}) and $2 u_- = G(z_1) + O(1/N)$.
This result with respect to the connected part agrees with that obtained by 
solving functional equations \cite{B,I} and by a diagrammatic method \cite{IS}.
Especially the disconnected part up to $1/N^2$ order in eq.~(\ref{GOE2})
is consistent with \cite{I}.

The saddle point evaluation holds only in the case of $1-2 u_- v_- = O(N^0)$,
as pointed out in the calculation of the one-point Green's function.
When we compute the following two level correlators: 
\beq
	\bra \frac{1}{N} \Tr \frac{1}{z_1 - H + i \ep} \, 
        \frac{1}{N} \Tr \frac{1}{z_2 - H - i \ep} \ket, 
\eeq
the coefficient of Gaussian fluctuations become $1-2 u_+ v_-$. 
The result of the saddle-point method 
 should be trusted only for $z_1 -z_2 = O(N^0)$
indicated by the definition of $u_\pm$ eq.~(\ref{upm}).
Therefore, the short-distance two-level correlators
should be calculated in other method instead of the saddle point evaluation.
\section{Gaussian unitary ensemble}
We turn to the case of the GUE, which means the ensemble of Hermitian 
matrices following 
the Gaussian distribution (\ref{prob}).  The measure $DH$ is explicitly
written as 
\beq
        D H \equiv \prod_{i \leq j} d \left( {\rm Re \, } H_{i  j} \right) 
        \prod_{i <  j} d \left( {\rm Im \, }  H_{i j} \right). 
\eeq
We can proceed computation along the way in  
the case of GOE.  We assume 
\beq
        | z | > 2
\label{assumptionh}
\eeq 
in the case of the GUE, which corresponds to the assumption 
(\ref{assumption}). 
We can construct a generating function $W(z)$  by employing 
{\it complex} vector variables as follows :
\beq
        W(z)  \equiv \bra \frac{1}{Z_\phi} \,  
        \int D  \phi \, \exp \{ -\frac{i}{2} {\phi^a }^\dagger \, 
        (z - H ) \, \phi^a  \} \ket,    
\label{Wh}
\eeq
where 
\beq
         D \phi \equiv \prod_{a = 1}^n \prod_{i =1}^N d 
         \left( {\rm Re \, } \phi_i^a \right) 
        d \left( {\rm Im \, } \phi_i^a \right).  
\eeq
We can readily show that 
\beq
        G(z) = \lim_{n \rightarrow 0} (-\frac{1}{n N}) \frac{d}{d z} W(z)
\label{WtoGh}
\eeq  
in a similar way to the derivation of (\ref{WtoG}).  
Difference between 
eq.~(\ref{WtoG}) and eq.~(\ref{WtoGh}) arises because 
the complex vector variables contribute twice comparing to the case 
of real vector variables. 
We first take the average over $H$ in (\ref{Wh}) and we introduce 
a Hermitian auxiliary matrix $Q_{a b}$ along the line with the case of 
the GOE. See eq.~(\ref{W(z)}) to eq.~(\ref{W(z)byQ}).   Then the
 integration over the complex vector variables gives  
\beqa
        W(z) &=& \frac{1}{Z_Q} \, \int DQ \, 
        \exp \left\{ -\frac{N}{2} \Tr 
		\left( 2 \, \log i(z - Q) + Q^2 \right) \right\} \nn\\
        &=&  \frac{C_n}{Z_Q} \int  \prod^{n}_{a=1} dw_a \ 
        {\Delta(w)}^2  \  \exp \{ -\frac{N}{2} 
\sum^{n}_{a = 1} \left( 2 \log i(z - w_a) 
        + {w_a}^2 \right) \}.
\label{Qh} 
\eeqa
Diagonalizing $Q$ and we get the saddle-point equation corresponding 
to eq.~(\ref{sp}) 
\beq
        - \frac{1}{z - w} +  w = 0.  
\eeq
The most dominant saddle-point is $w_- I_n$, in which 
\beq
        w_- \equiv \frac{1}{2} (z - \sqrt{z^2 - 4}). 
\label{w-} 
\eeq
 We expand the exponent of (\ref{Qh}) around $w_- I_n$ and 
ignore  the cubic and higher terms of the fluctuation as in the same way
 of GOE.  
The Gaussian integration of the fluctuation is easily performed, from 
which we derive 
\beq
        W(z) \sim \exp \left\{ - \frac{N n}{2} 
        \left( 2 \log(z - w_-) + w_-^2 \right)\right\} 
        {\left( \frac{1}{1 - w_-^2} \right)}^{n^2/2}
\eeq
up to the next-to-leading order.  Using this result and 
eq.~(\ref{WtoGh}), we get 
\beq
        G(z) = \frac{1}{2}(z - \sqrt{z^2 -4}) + O(1/N^2).   
\eeq
Note that  first-order corrections vanish in this case.

The two-point function defined in eq.~(\ref{G2}) is also calculated
 following the case of GOE.  Using complex vectors, we define the
 generating function $W(z_1, z_2)$ as 
\beq
        W(z_1, z_2) \equiv \bra  \frac{1}{Z_{\phi_1}Z_{\phi_2}}  
        \int D\phi_1 D\phi_2 \, 
         \exp \{ -\frac{i}{2} \,  {\phi^a_{1}}^\dagger \, (z_1 \,  - H) \, 
        \phi^a_{1} -\frac{i}{2} \,  {\phi^p_{2}}^\dagger \, (z_2 \,  - H ) \, 
        \phi^p_{2}  \} \ket.
\eeq
The two-point function $G(z_1, z_2)$ is derived from $W(z_1, z_2)$ as 
\beq
        G(z_1, z_2) = \lim_{n, m \rightarrow 0} 
        \left( \frac{-1}{Nm}\right) \left(\frac{-1}{Nn}\right) 
\frac{\partial ^2}{\partial z_1 \partial z_2} W(z_1, \, z_2). 
\label{W2toG2h}
\eeq
We can write  $W(z_1, z_2)$ in terms of an integration over 
Hermitian $(n+m) \times (n+m)$ matrices, 
which corresponds to eq~(\ref{W2byQ}) for the GOE case.  
 The result is 
\beq
                W(z_1, \, z_2) = \frac{1}{Z_Q} \int DQ \, 
                \exp \left \{-\frac{N}{2} \Tr  \left(2 \log i(\bz - Q) +  
				Q^2 \right)
                \right \}. 
\eeq
The most dominant saddle point $\bar Q$ is 
\beq
        \bar Q = 
        \left(
        \begin{array}{cc}
        w_1 I_n & 0 \\
                0 & w_2 I_m \\
        \end{array}
        \right),   
\label{barQh}
\eeq
where 
\beq
        w_i \equiv \frac{1}{2} \left(z_i - \sqrt{{z_i}^2 - 4} \right), \qquad 
        i = 1,2.
\eeq
Since the connected part of the two-point function is the order 
$O(1/N^2)$, 
we must take in the effect of fluctuations around $\bar Q$.  As we
 explained 
in the GOE case, we can regard the fluctuations as Gaussian for the 
leading order 
of the connected part.  Following the step to derive eq.~(\ref{W2-2}), 
we get 
\beq
        W(z_1, z_2) = W(z_1) W(z_2) \left( \frac{1}{1 - w_1 w_2} \right)^{nm}.  
\eeq
Employing eq.~(\ref{W2toG2h}), the above result is  translated to the
 language of the Green's function : 
\beq
        G(z_1, z_2) = G(z_1) G(z_2) - \frac{1}{N^2} \frac{\del^2}{\del z_1 
        \del z_2} 
        \log \left(1 - G(z_1) G(z_2) \right).  
\eeq
This agrees with the well-known result obtained by several other methods
\cite{AJM,BZ,B,I,IS}. 

\section{Gaussian symplectic ensemble}
\subsection{Definition}
The Gaussian symplectic ensemble(GSE) is the ensemble  of 
 quaternion real Hermitian matrices with the distribution  
(\ref{prob}).  
We shall first recall the definition of a quaternion  real Hermitian
 matrix. 

Let us write 
\beq
	H =  \left(
	       \begin{array}{ccc}
	        H_{11} & \ldots & H_{1N} \\
	        H_{21} & \ldots  & H_{2 N} \\
	        \multicolumn{3}{c}{\dotfill}\\
	        H_{N1} & \ldots & H_{N N} \\
	       \end{array}
	       \right), 
\eeq 
where each entry $H_{ij}$ is a quaternion number, which can be
 represented as a $2 \times 2$ 
matrix. We choose the set of the Pauli matrices $\sigma^k$ and the 
unit matrix $I_2$ as a basis of $2 \times 2$ matrices.  
Then $H_{i j}$ is written as 
\beqa
	H_{i j} &=& H^{(0)}_{i j} I_2 + i \sum_{k = 1}^3 H^{(k)}_{i j} 
                    	\sigma^k \nn \\
	           &\equiv& 
           \left(
	       \begin{array}{cc}
	        H_{ij}^{11} &  H_{ij}^{12} \\
	        H_{ij}^{21} &  H_{ij}^{22} \\
	       \end{array}
	       \right). 
\eeqa
In our notation, $i, \, j =1, \cdots,  N $  and 
 $ \al, \, \be = 1, \,  2$ in $H_{ij}^{\al\be}$.
The matrix $H$ is called a quaternion real Hermitian matrix if 
and only if $H^{(0)}_{i j}$ forms a real symmetric matrix whereas 
$H^{(k)}_{i j}$ $(k = 1,2,3)$ form real antisymmetric matrices. 
These conditions are alternatively expressed as 
\beqa
	 &&{H_{i j}^{11}}^* =  {H_{i j}^{22}} \nn\\
	 &&{H_{i j}^{12}}^* = - {H_{i j}^{21}}\nn\\
	 &&{H_{i j}^{\dagger}} = H_{j i}. 
\label{qrh}
\eeqa
Note that ${}^{\dagger}$ in the last equation means the Hermitian
conjugation when we regard $H_{i j}$ as a $2 \times 2$ matrix. 

The trace of $H^2$ in eq.(\ref{prob}) reads 
\beq
	\Tr H^2 = {H_{i j}^{\al \be}}{H_{ji}^{\be \al}}, 
\eeq
where $\al, \be = 1,  2$.  
Using $H^{(\mu)}_{i j}$,  $\mu = 1, \cdots 4$, 
the measure $D H$ in this case is written as 
\beq
	D H  = \prod_{i \leq j} d H_{ij}^{(0)}
	\prod_{k = 1}^3  \prod_{i < j} d H_{ij}^{(k)}. 
\eeq

Here, for later convenience,  we consider how to construct 
a quaternion real Hermitian matrix from 
a complex vector with an even-number component.  
Let $\phi_i^{a \al}$ ($1 \leq i \leq N$,  $ 1 \leq a \leq n$, 
$\al = 1,2$ ) be a complex number.  We regard 
$\phi^{a}$'s as the $2 n$-component vectors. 
The diadig  
\beq
	\tilde \ka(\phi^a) \equiv 
	 \phi^{a} {\phi^{a}}^{\dagger}  \qquad
	{\rm or } \qquad \tilde \ka_{ji}^{\al \be} \equiv 
	\phi^{a \be}_{j} {\phi^{a \al}_{i}}^{*}  
\eeq 
defines the  $2 N \times 2 N$ Hermitian matrix. 
For constructing a quaternion real Hermitian matrix, 
we need to ``symmetrize'' $\tilde \ka(\phi^a)$  in the following way: 
\beqa
	\tilde \ka^{1 1}_{j i} &\rightarrow& \frac{1}{2} \left( 
	\tilde  \ka^{1 1}_{j i} + \tilde  \ka^{2 2}_{i j} \right)  
	\equiv \ka^{1 1}_{j i} \nn\\
	\tilde \ka^{1 2}_{j i} &\rightarrow& \frac{1}{2} \left( 
	\tilde \ka^{1 2}_{j i} - \tilde \ka^{1 2}_{i j} \right) \equiv 
	\ka^{1 2}_{j i}\nn\\
	\tilde \ka^{2 1}_{j i} &\rightarrow& \frac{1}{2} \left( 
	\tilde  \ka^{2 1}_{j i} - \tilde  \ka^{2 1}_{i j} \right) 
	\equiv \ka^{2 1}_{j i}\nn\\
	\tilde \ka^{2 2}_{j i} &\rightarrow& \frac{1}{2} \left( 
	\tilde  \ka^{2 2}_{j i} + \tilde  \ka^{1 1}_{i j} \right) 
	\equiv \ka^{2 2}_{j i}. 
\label{symmetrize}
\eeqa
The matrix $\ka(\phi^{a})$ defined above 
is a quaternion real Hermitian matrix since it satisfies the condition 
\beqa
	 &&{\ka_{i j}^{11}}^* =  {\ka_{i j}^{22}} \nn\\
	 &&{\ka_{i j}^{12}}^* = - {\ka_{i j}^{21}}\nn\\
	 &&{\ka_{i j}^{\dagger}} = \ka_{j i}. 
\eeqa

Another useful quaternion real Hermitian matrix is constructed 
as follows: 
we  define 
\beq
	{\psi_i}^{a 1} \equiv {\phi_i}^{a 1} , \qquad 
	{\psi_i}^{a 2} \equiv {{\phi_i}^{a 2}}^{*},   
\label{psi}
\eeq
and make the diadig $\psi_i \psi_i^\dagger$.  
By the same symmetrization as eq.~(\ref{symmetrize}), 
we get the  $2 n \times 2n$ quaternion real Hermitian matrix.  
We shall denote the resultant matrix by 
$\la(\phi_i)$.  A straightforward calculation gives 
\beq
	\Tr \, \ka(\phi^a)^2 = \Tr \, \la(\phi_i)^2.  
\label{kandl}
\eeq

\subsection{One-point function}
Let us compute the one-point function $G(z)$ averaged over GSE 
\beq
	     G(z) \equiv \langle \frac{1}{2 N} \Tr \frac{1}{z - H} \rangle. 
\eeq 
Since $H$ is a $2 N \times 2 N$ matrix, $G(z)$ tends to $1/z$ as 
$z \rightarrow \infty$.  We assume (\ref{assumptionh}) as in the 
case of the GUE.  The generating function $W(z)$ 
is defined by 
\beq
	     W(z)  \equiv \bra \frac{1}{Z_\phi} \,  
        \int D  \phi \, \exp \{ -\frac{i}{2} {\phi^a }^\dagger \, 
        (z - H ) \, \phi^a  \} \ket. 
\label{Ws}    
\eeq
Note that $\phi^a$ is a $2 N$-component vector, which implies 
\beq
	{\phi^a }^\dagger \,  H  \, \phi^a \equiv 
	{\phi^{a \al}_{i}}^* \,  H_{i j}^{\al \be} \, \phi^{a \be}_j 
\label{pHp} 
\eeq
in eq.(\ref{Ws}).  
The one-point function is derived from $W(z)$ as 
\beq
        G(z) = \lim_{n \rightarrow 0} (-\frac{1}{2 n N}) 
	     \frac{d}{d z} W(z). 
\label{WtoGs}
\eeq  

Let us take the average over $H$ in eq.~(\ref{Ws}).  
We first notice that 
\beq
	{\phi^a }^\dagger \,  H  \, \phi^a =  \Tr \, \tilde \ka(\phi^a) H. 
\label{coupling}
\eeq
We can symmetrize $\tilde \ka (\phi^a)$ using the following 
relations

\beq	
 H^{11}_{i j} = H^{22}_{ji}, \qquad H^{12}_{i j} = - H^{12}_{ji}. 
\eeq 
  which are derived from eqs.~(\ref{qrh}).  
For example, the coupling with the diagonal part of 
$H_{i j}$ in eq.~(\ref{coupling})  becomes 
\beqa
 \tilde \ka^{11}_{j i} H_{i j}^{11} +  \tilde \ka^{22}_{j i} H_{i j}^{22}
&=& 
\frac{1}{2} \left( \tilde \ka^{11}_{j i} + \tilde \ka^{22}_{ij} \right) 
H_{i j}^{11} + \frac{1}{2} \left( \tilde \ka^{11}_{ij} + \tilde \ka^{22}_{ji} \right) H_{i j}^{22} \nn\\ 
&=& \ka^{11}_{j i} H_{i j}^{11} + \ka^{22}_{j i} H_{i j}^{22}
\eeqa
After a similar procedure is performed to 
 $H^{12}_{i j}$ and $H^{21}_{i j}$, it turns out that the matrix  
$\ka (\phi^a)$ defined in eq.~(\ref{symmetrize})
can replace $\tilde \ka(\phi^a)$ in eq.~(\ref{coupling}).  

We are ready to take the average over $H$ 
in eq~.(\ref{Ws}).  That is, 
\beqa
	- \frac{N}{2} \Tr \, H^2 + \frac{i}{2} {\phi^a }^\dagger \, H  \, \phi^a 
	= - \frac{N}{2} \Tr \, (H - \frac{i}{2 N} \ka(\phi^a))^2 - \frac{1}{8 N} 
	\Tr \, \ka(\phi^a)^2
\eeqa
The change  $H \rightarrow H + \frac{i}{2 N} \ka(\phi^a)$ 
can be performed by
 shifting the integration variables $H_{ij}^{(\mu)}$ 
because $\ka(\phi^a)$ is 
a quaternion real Hermitian matrix.   It leads to 
\beq
	W(z) = \frac{1}{Z_\phi} \,  \int D  \phi \, \exp \left(-\frac{i}{2} 
	{\phi^a }^\dagger \, z  \, \phi^a  - \frac{1}{8 N} 
	\Tr \, \ka(\phi^a)^2 \right).  
\label{byk}
\eeq

Next, for the sake of integration over $\phi$, 
we introduce the auxiliary matrix $Q$ that is a $2 n \times 2n$
quaternion real Hermitian matrix.
Let us consider the following integration 
\beq
	\frac{1}{Z_Q} \int D Q \, \exp \left ( - \frac{N}{2} Q_{a b}^{\al \be} 
	Q_{b a}^{\be \al} + 
	\frac{i}{2}  {\psi_i^{a \al} }^* Q_{a b}^{\al \be} 
	\psi^{b \be}_i \right), 
\label{auxs} 
\eeq 
where $\psi_i$'s $( i = 1, \cdots n)$ are defined in eq.~(\ref{psi}). 

The $Q$-integral in eq.~(\ref{auxs}) is carried out in the same way as
 the $H$-integral of eq.~(\ref{Ws}).  The result is  
\beq
	\frac{1}{Z_Q} \int D Q \, \exp \left ( - \frac{N}{2} Q_{a b}^{\al \be} 
	Q_{b a}^{\be \al} + 
	\frac{i}{2}  {\psi_i^{a \al} }^* Q_{a b}^{\al \be} 
	\psi^{b \be}_i \right) = \exp \left( - \frac{1}{8 N} 
	\Tr \, \la(\phi^a)^2 \right).  
\label{byQ}
\eeq
Thus, from eqs.~(\ref{psi}),(\ref{kandl}),(\ref{byk})   
and (\ref{byQ}), we conclude 
\beq
	W(z) = \frac{1}{Z_\phi Z_Q} \int DQ D\phi \, 
	\left ( - \frac{N}{2} \Tr \, Q^2 -  
	\frac{i}{2}  {\psi_i}^\dagger (z - Q) {\psi_i} \right).  
\eeq
Performing the integration over $\phi$, 
we get 
\beq
	W(z) = \frac{1}{Z_Q} \, \int DQ \, 
        \exp \left\{ -\frac{N}{2} \Tr \, 
		\left( 2 \, \log i(z - Q) + Q^2 \right) \right\}.  
\eeq
Since any eigenvalue of $Q$ has two-fold degeneracy \cite{mehta}, the
 eigenvalue representation of the above integration becomes 
\beq
	W(z) = \frac{1}{Z_Q} \int \, \prod_{a=1}^n dw_a \, {\Delta(w)}^4
        \exp \left\{ -N \sum_{a=1}^n 
		\left( 2 \, \log i(z - w_a) + {w_a}^2 \right) \right\}
\eeq

Repeating the same argument as in the case of GUE, we obtain the 
 result up to the next-to leading order:
\beq
        W(z) \sim \exp \left\{ - N n 
        \left( 2 \log(z - w_-) + w_-^2 \right)\right\} 
        {\left( \frac{1}{1 - w_-^2} \right)}^{n^2-n/2}, 
\eeq
where $w_-$ is defined in eq.~(\ref{w-}).  The exponent of $(1 - w_-^2)$
 is different from the result of GUE because the degree of freedom of
 fluctuations in the case of the GUE is given by $n^2$ while $2 n^2 -n$ in 
the case of the GSE. That affects the non-vanishing first-order 
correction to $G(z)$. 
Namely, from eq.~(\ref{WtoGs}), we obtain 
\beq
	G(z) = \frac{1}{2} 
	\left( z - \sqrt{z^2 - 4} \right)
	\left(1 - \frac{1}{2 N}\frac{1}{z^2 -4} \right ) + O(1/N^2).  
\eeq

\subsection{Two-point function}
Next we compute the two-point function $G(z_1, z_2)$ defined 
by  
\beq
	G(z_1, z_2) \equiv \bra \frac{1}{2 N} \Tr \frac{1}{z_1 - H} \, 
        \frac{1}{2 N} \Tr \frac{1}{z_2 - H} \ket.
\eeq
 The generating function $W(z_1, z_2)$ has the same form as 
the GUE case: 
\beq
        W(z_1, z_2) \equiv \bra  \frac{1}{Z_{\phi_1}Z_{\phi_2}}  
        \int D\phi_1 D\phi_2 \, 
         \exp \{ -\frac{i}{2} \,  {\phi^a_{1}}^\dagger \, (z_1 \,  - H) \, 
        \phi^a_{1} -\frac{i}{2} \,  {\phi^p_{2}}^\dagger \, (z_2 \,  - H ) \, 
        \phi^p_{2}  \} \ket.
\eeq
The two-point function $G(z_1, z_2)$ is 
related to  $W(z_1, z_2)$ as follows:
\beq
        G(z_1, z_2) = \lim_{n, m \rightarrow 0} 
        \left( \frac{-1}{2Nm}\right) \left(\frac{-1}{2Nn}\right) 
	\frac{\partial ^2}{\partial z_1 \partial z_2}    W(z_1, \, z_2). 
\label{W2toG2s}
\eeq

In order to perform the integration over $H$, we introduce the 
notation $\Phi^{A \al}_i$ as 
\beq
        \Phi^{A \, \al}_i \equiv 
	\left \{
       \begin{array}{ll}
	   \phi^{A \, \al}_{1 i} & 1 \leq A \leq n \\
       \phi^{A-n \, \al}_{2 i} & n+1 \leq A \leq n+m
       \end{array}
	   \right..    
\eeq 
Then, since the coupling to $H$ can be rewritten as 
${\Phi^A}^\dagger H \Phi^A$,  we get the following result as 
the integration over $H$: 
\beq
	 W(z_1, \, z_2) = \frac{1}{Z_{\phi_1}Z_{\phi_2}}  
        \int D\phi_1 D\phi_2 \, 
         \exp \{ -\frac{i}{2} \,  {\Phi^A}^\dagger \, \bz \, \Phi^A - 
	     \frac{1}{8 N} \Tr \, \ka(\Phi^A)^2 \}, 
\eeq
where $\bz$ is the following $2(n+m) \times 2(n+m)$ matrix 
\beq
        \bz = 
        \left(
        \begin{array}{cc}
        z_1 I_{2 n} & 0 \\
                0 & z_1 I_{2 m}\\
        \end{array}
        \right).
\label{barQq}
\eeq
According to the formulae (\ref{kandl}) and (\ref{byQ}), we can 
express $W(z_1 \, z_2)$ by $2(n+m) \times 2(n+m)$ matrix $Q$: 
\beqa
	W(z_1, \, z_2) &=& \frac{1}{Z_{\phi_1}Z_{\phi_2}Z_Q}  
     \int D\phi_1 D\phi_2 DQ \, 
     \exp \{ -\frac{i}{2} \,  {\Phi_i}^\dagger \, \bz \, \Phi_i - 
	 \frac{N}{2} \Tr \, Q^2 +  
	\frac{i}{2}  {\Psi_i}^\dagger  Q {\Psi_i}  \} \nn\\
	&=& \frac{1}{Z_{\phi_1}Z_{\phi_2}Z_Q}  
     \int D\phi_1 D\phi_2 DQ \, 
     \exp \{ - \frac{N}{2} \Tr \, Q^2 -  
	 \frac{i}{2}  {\Psi_i}^\dagger \left(z- Q \right) {\Psi_i}  \}, 
\eeqa
where $\Psi$ is defined in the same way as eq.~(\ref{psi}): 
\beq
	{\Psi_i}^{A 1} \equiv {\phi_i}^{A 1} , \qquad 
	{\Psi_i}^{A 2} \equiv {{\phi_i}^{A 2}}^{*},   
\label{Psi}
\eeq
After the integration over the vector variables, we get 
\beq
	W(z_1, \, z_2) = \frac{1}{Z_Q} \int DQ \, 
	\exp \left \{-\frac{N}{2} \Tr  \left(2 \log i(\bz - Q) +  
	Q^2 \right)
	\right \}. 
\eeq
The most dominant saddle point $\bar Q$ is 
\beq
        \bar Q = 
        \left(
        \begin{array}{cc}
        w_1 I_{2 n} & 0 \\
        0 & w_2 I_{2 m} \\
        \end{array}
        \right).   
\label{barQs}
\eeq

Next we compute effects of fluctuations around $\bar Q$. 

\beq
	W(z_1, z_2) = W(z_1) W(z_2) \left( \frac{1}{1 - w_1 w_2} 
	\right)^{2 nm}.  
\eeq
Employing eq.~(\ref{W2toG2h}), the above result is  translated to the
 language of the Green's function : 
\beq
	G(z_1, z_2) = G(z_1) G(z_2) - \frac{1}{2 N^2} 
	\frac{\del^2}{\del z_1 
	\del z_2} 
	\log \left(1 - G(z_1) G(z_2) \right).
\label{GSE2}  
\eeq
This result with respect to the connected part agrees with that obtained by 
solving functional equations \cite{B,I} and by a diagrammatic method \cite{IS}.
Especially the disconnected part up to $1/N^2$ order in eq.~(\ref{GSE2})
is consistent with \cite{I}. 

\section{Three-point function}
Finally, we calculate the connected three-point function 
in each Gaussian ensemble.
That in the GOE case is defined by
\begin{eqnarray}
G(z_1,z_2,z_3) \equiv \langle
\frac{1}{N} \Tr \frac{1}{z_1-H}
\frac{1}{N} \Tr \frac{1}{z_2-H}
\frac{1}{N} \Tr \frac{1}{z_3-H}   \rangle.
\end{eqnarray}
As is the case of one and two-point function, 
we construct the generating function of $G(z_1,z_2,z_3)$
using three $n$-flavor {\it real} vectors:
\begin{eqnarray}
W(z_1,z_2,z_3)=
\langle
\frac{1}{Z_{\phi_1}Z_{\phi_2}Z_{\phi_3}}
\int \prod_{i=1}^{3}D{\phi}_{i}
\exp \{ -\sum_{j=1}^{3}
\frac{i}{2}\trans\phi_{j}^{a}(z_{j}-H)\phi_{j}^{a} \}
\rangle.
\end{eqnarray}
The three-point function $G(z_1,z_2,z_3)$ 
is derived from $W(z_1,z_2,z_3)$ as
\begin{eqnarray}
G(z_1,z_2,z_3)=\lim_{n \to 0}(\frac{-2}{Nn})^3
\frac{\partial^3}{\partial z_1 \partial z_2 \partial z_3}
W(z_1,z_2,z_3).
\label{WtoG3O}
\end{eqnarray}
We can write $W(z_1,z_2,z_3)$ in the term of an integration over $3n \times 3n$ real symmetric matrix. The result is
\begin{eqnarray}
W(z_1,z_2,z_3)=\frac{1}{Z_Q} \int DQ
\exp \{ -\frac{N}{2}\Tr(\log i({\bf{z}}-Q)+Q^2)\}.
\end{eqnarray}
The most dominant saddle point $\bar{Q}$ is
\begin{eqnarray}
\bar{Q}=\left(
 \begin{array}{ccc}
  w_1 I_n & 0 & 0 \\
  0 & w_2 I_n & 0 \\
  0 & 0 & w_3 I_n 
 \end{array}
\right),  
\end{eqnarray}
where
\begin{eqnarray}
w_i \equiv \frac{1}{2}(z_i-\sqrt{z_i{}^2-2}),
\quad\quad
i=1,2,3.
\end{eqnarray}

Next we consider fluctuations around the saddle point
$\bar{Q}$.
\begin{eqnarray}
W(z_1,z_2,z_3)=\frac{e^{-S_0}}{Z_Q} \int D(\delta Q)
e^{-S_2-\sum_{l=3}^{\infty}S_l},
\end{eqnarray}
where
\begin{eqnarray}
S_0&=&\frac{N}{2}\Tr(\log i({\bf{z}}-\bar{Q})+\bar{Q}^2), \\
S_2&=&\frac{1}{2}
      \Tr((\delta Q)^2-(\bar{Q}\delta Q)^2), \nonumber \\
   &=&\frac{1}{2}\sum_{i,j=1}^{3}\sum_{a,b=1}^{n}
      (1-w_{i}w_{j})
      (\delta Q_{ab}^{ij})^2, \\
S_l&=&\frac{1}{lN^{l/2-1}}\Tr(\bar{Q}\delta Q)^l,
      \quad\quad l>3.
\end{eqnarray} 
Since the connected part of three-point function is in the order
$1/N^4$, we must calculate the following Gaussian integral with $S_4$
and $S_3^2$ terms:
\begin{eqnarray}
W(z_1,z_2,z_3)&=&\frac{e^{-S_0}}{Z_Q} \int D(\delta Q)
e^{-S_2}\{1-\sum_{l=3}^{\infty}S_l
+\frac{1}{2!}(\sum_{l=3}^{\infty}S_l)^2-\cdots\},\nonumber \\
&=&e^{-S_0}\{1-
\langle\langle S_4 \rangle\rangle
+\frac{1}{2!}\langle\langle S_3^2 \rangle\rangle +O(1/N^5)\},
\label{2terms}
\end{eqnarray}
where
\begin{eqnarray}
\langle\langle\cdots\rangle\rangle
=\frac{1}{Z_Q} \int D(\delta Q)(\cdots)e^{-S_2}.
\end{eqnarray}
Calculating eq.(\ref{2terms}), we have
\begin{eqnarray}
W(z_1,z_2,z_3)=\frac{n^3}{2N}F(z_1,z_2,z_3),
\end{eqnarray}
where
\begin{eqnarray}
&&F(z_1,z_2,z_3)
=\frac{X_{12}}{1-X_{12}}\frac{X_{23}}{1-X_{23}}\frac{X_{31}}{1-X_{31}} \nonumber \\
&&
+\frac{X_{31}}{1-X_{31}}\frac{X_{12}}{1-X_{12}}\frac{1}{1-X_{11}}
+\frac{X_{12}}{1-X_{12}}\frac{X_{23}}{1-X_{23}}\frac{1}{1-X_{22}}
+\frac{X_{23}}{1-X_{23}}\frac{X_{31}}{1-X_{31}}\frac{1}{1-X_{33}},
\end{eqnarray}
\begin{eqnarray}
X_{ij}
=2w_{i}w_{j}
=\frac{1}{2}G(z_{i})G(z_{j}).
\end{eqnarray}
Substituting the above formula into eq.(\ref{WtoG3O}),
we can derive the three-point function
\begin{eqnarray}
G_C(z_1,z_2,z_3)=-\frac{4}{N^4}
\frac{\partial^3}{\partial z_1 \partial z_2 \partial z_3}
F(z_1,z_2,z_3).
\end{eqnarray}
The connected part in this result
agrees with that obtained by other methods \cite{VWZ,I,IS}.

Next we turn to the connected three-point function in the GUE case
along the way in the case of GOE.
We construct a generating function $W(z_1,z_2,z_3)$ by employing
three {\it{complex}} vectors as follows:
\begin{eqnarray}
W(z_1,z_2,z_3)=\langle
\frac{1}{Z_{\phi_1}Z_{\phi_2}Z_{\phi_3}}
\int \prod_{i=1}^{3}D{\phi}_{i}
\exp \{ -\sum_{j=1}^{3}
\frac{i}{2}
\phi_{j}^{a}{}^{\dagger}(z_{j}-H)
\phi_{j}^{a} \}
\rangle.
\end{eqnarray}
The three-point function is derived from $W(z_1,z_2,z_3)$ in the following
formula:
\begin{eqnarray}
G(z_1,z_2,z_3)=\lim_{n \to 0}(\frac{-1}{Nn})^3
\frac{\partial^3}{\partial z_1 \partial z_2 \partial z_3}
W(z_1,z_2,z_3).
\label{WtoG3U}
\end{eqnarray}
$W(z_1,z_2,z_3)$ is written in the form of $Q$-integral
\begin{eqnarray}
W(z_1,z_2,z_3)=\frac{1}{Z_Q} \int DQ
\exp \{ -\frac{N}{2}\Tr(2 \log i({\bf{z}}-Q)+Q^2)\},
\end{eqnarray}
where $Q$ means $3n{\times}3n$ hermitian matrix.
 The most dominant saddle point $\bar{Q}$ is
\begin{eqnarray}
\bar{Q}=\left(
 \begin{array}{ccc}
  w_1 I_n & 0 & 0 \\
  0 & w_2 I_n & 0 \\
  0 & 0 & w_3 I_n 
 \end{array}
\right),  
\end{eqnarray}
where
\begin{eqnarray}
w_i \equiv \frac{1}{2}(z_i-\sqrt{z_i{}^2-4}),
\quad\quad
i=1,2,3.
\end{eqnarray}
The dominant contributions to $W(z_1,z_2,z_3)$ becomes
\begin{eqnarray}
W(z_1,z_2,z_3)=\frac{n^3}{N}F(z_1,z_2,z_3),
\end{eqnarray}
where
\begin{eqnarray}
&&F(z_1,z_2,z_3)
=\frac{X_{12}}{1-X_{12}}\frac{X_{23}}{1-X_{23}}\frac{X_{31}}{1-X_{31}} \nonumber \\
&&
+\frac{X_{31}}{1-X_{31}}\frac{X_{12}}{1-X_{12}}\frac{1}{1-X_{11}}
+\frac{X_{12}}{1-X_{12}}\frac{X_{23}}{1-X_{23}}\frac{1}{1-X_{22}}
+\frac{X_{23}}{1-X_{23}}\frac{X_{31}}{1-X_{31}}\frac{1}{1-X_{33}},
\end{eqnarray}
\begin{eqnarray}
X_{ij}=w_{i}w_{j}=G(z_i)G(z_j).
\end{eqnarray}
Thus, the three-point function in GUE is
\begin{eqnarray}
G_C(z_1,z_2,z_3)=-\frac{1}{N^4}
\frac{\partial^3}{\partial z_1 \partial z_2 \partial z_3}
F(z_1,z_2,z_3).
\end{eqnarray}
The connected part in this result
agrees with that obtained by other methods \cite{AJM,VWZ,I,IS}.

To end with, we calculate the connected three-point function in GSE case.
That in this case is defined by 
\begin{eqnarray}
G(z_1,z_2,z_3) \equiv \langle
\frac{1}{2N} \Tr \frac{1}{z_1-H}
\frac{1}{2N} \Tr \frac{1}{z_2-H}
\frac{1}{2N} \Tr \frac{1}{z_3-H}   \rangle.
\end{eqnarray}
The three-point function derived from generating function is as follows:
\begin{eqnarray}
G(z_1,z_2,z_3)=\lim_{n \to 0}(\frac{-1}{2Nn})^3
\frac{\partial^3}{\partial z_1 \partial z_2 \partial z_3}
W(z_1,z_2,z_3).
\label{WtoG3S}
\end{eqnarray}
And the generating function is calculated as
\begin{eqnarray}
W(z_1,z_2,z_3)=\frac{1}{Z_Q} \int DQ
\exp \{ -\frac{N}{2}\Tr(2 \log i({\bf{z}}-Q)+Q^2)\},
\end{eqnarray}
where $Q$ means $6n{\times}6n$ quaternion real hermitian matrix.
The most dominant saddle point $\bar{Q}$ is
\begin{eqnarray}
\bar{Q}=\left(
 \begin{array}{ccc}
  w_1 I_{2 n} & 0 & 0 \\
  0 & w_2 I_{2 n} & 0 \\
  0 & 0 & w_3 I_{2 n} 
 \end{array}
\right),  
\end{eqnarray}
where
\begin{eqnarray}
w_i \equiv \frac{1}{2}(z_i-\sqrt{z_i{}^2-4}),
\quad\quad
i=1,2,3.
\end{eqnarray}
Performing $1/N$-expansion for $W(z_1,z_2,z_3)$,
\begin{eqnarray}
W(z_1,z_2,z_3)=\frac{4n^3}{N}F(z_1,z_2,z_3)
\end{eqnarray}
where
\begin{eqnarray}
&&F(z_1,z_2,z_3)
=\frac{4X_{12}}{1-X_{12}}\frac{X_{23}}{1-X_{23}}\frac{X_{31}}{1-X_{31}} \nonumber \\
&&
+\frac{X_{31}}{1-X_{31}}\frac{X_{12}}{1-X_{12}}\frac{1}{1-X_{11}}
+\frac{X_{12}}{1-X_{12}}\frac{X_{23}}{1-X_{23}}\frac{1}{1-X_{22}}
+\frac{X_{23}}{1-X_{23}}\frac{X_{31}}{1-X_{31}}\frac{1}{1-X_{33}},
\end{eqnarray}
\begin{eqnarray}
X_{ij}=w_{i}w_{j}=G(z_i)G(z_j).
\end{eqnarray}
The three-point function becomes
\begin{eqnarray}
G_C(z_1,z_2,z_3)=-\frac{1}{2N^4}
\frac{\partial^3}{\partial z_1 \partial z_2 \partial z_3}
F(z_1,z_2,z_3).
\end{eqnarray}
The connected part in this result
agrees with that obtained by other methods \cite{I,IS}.

\section{Concluding remarks}

We have calculated the averaged one point Green's functions, the wide
connected two-level and three-level correlators 
in Gaussian orthogonal, unitary and
symplectic random matrix ensembles by the replica method. 
The one-point Green's functions have been 
calculated to the next-to-leading order
in the $1/N$ expansion in GOE and GSE. 
Our results are consistent with those by other methods \cite{B,I,IS}.
 We have notified that there are some regions 
of the spectral parameter $z$ where the employed saddle point evaluation cannot
work well for averaged Green's functions. 
In those regions, the fluctuation of 
the auxiliary variable $Q$  becomes large and the 
higher orders in the saddle point
expansion give the same contribution with the order $O(N^0)$.   
To calculate the short distance correlator we have to employ 
other method for the integration instead of the saddle point evaluation. \\



\newpage

\end{document}